# Software for Wearable Devices: Challenges and Opportunities

He Jiang[1], Xin Chen[1], Shuwei Zhang[1], Xin Zhang[1], Weiqiang Kong[1], Tao Zhang[2]

[1]School of Software, Dalian University of Technology, Dalian, China
[2]Department of Computing, The Hong Kong Polytechnic University, Kowloon, Hong Kong
jianghe@dlut.edu.cn

*Abstract*—Wearable devices are a new form of mobile computer system that provides exclusive and user-personalized services. Wearable devices bring new issues and challenges to computer science and technology. This paper summarizes the development process, current situation and important software research issues of wearable devices.

## I. INTRODUCTION

### A. The Definition of Wearable Devices

A wearable device is a computer that is subsumed into the personal space of a user, controlled by the user, and has both operational and interactional constancy, i.e., is always on and always accessible [1]. Wearable devices have the same computing abilities as mobile phones and tablet computers. In some cases, however, wearable devices are more competent for tasks such as calculation, navigation, remote picture than handheld devices due to their portability and characteristics to be detailed below.

### B. The Development of Wearable Devices

We can have a clear understanding of the development for wearable devices from [2][3]. Wearable devices have undergone many years of development since the initial ideas and prototypes appeared in the 1960s. During the 1960s to 1970s, wearable devices were in their embryonic period. People designed wearable devices for special purposes, interests or events. During the period, wearable devices remained in a small-scale field and people rarely understood their roles. In 1966, Edward Thorp, a professor in the Massachusetts Institute of Technology (MIT), invented a pair of shoes that could be used to cheat at roulette. This is the first wearable device in the world. In 1975, the Hamilton Watch Company launched a "calculator" watch which is the world's first wrist calculator. In 1977, the CC Collins designed a wearable device for the blind, which converts images captured from a head-mounted camera into tactile grids located on the blind's vests.

During the 1980s to 1990s, wearable devices entered the primary stage of development. People began to pay attention to wearable devices. Although wearable technology had a great improvement, wearable devices were still not practical for consumers and not friendly for users. In 1981, Steve Mann designed a head-mounted camera that to some extent can be regarded as the pioneer of Google glasses. In the same year, Steve Mann designed a backpack style computer with text, image and multimedia functions, displaying through the helmet. In 1997, Massachusetts Institute of Technology, Carnegie Mellon University, and Georgia Institute of Technology jointly organized the first International Symposium on Intelligent Wearable Computer (ISWC). Since then, smart wearable computing and smart wearable devices have attracted wide attention in academia and industry.

Since the 21st century, wearable devices have entered an advanced stage of development and aroused widespread concern. They become more complex and are designed according to the needs of users or the market. Many companies launched independently designed wearable devices and released corresponding software and hardware development platforms. In 2007, James Park and Eric Friedman founded the Fitbit Company that is dedicated to the development of wearable devices on pedometer and sleep quality detection etc. In 2013, Google launched Google glass and caused a sensation in the world. Meanwhile, Apple, Samsung, Sony and other companies have been developing their smart watches.

In the next few years, predictably, wearable devices will enter a period of prosperity. The IMS data [4] revealed that wearable devices shipments will reach 92.5 million units by 2016. According to Juniper's research [5], the number of wearable devices including smart watches and glasses will approach 130 million by 2018. Moreover, according to IDC's reports [6], we note that wearable devices had been under a rapid development; the number should be up to 19 million by the end of 2014; and the predicted shipments including smart watches and related devices will grow at an annual rate of 78% and reach 112 million by 2018. Therefore, we can believe that wearable devices will gradually enter people's lives and bring convenience to human, and wearable market will attract more participants.

### C. Classification Standards for Wearable Devices

At present, there are two standards for classifying wearable devices [7]. One standard is based on product forms, including head-mounted (such as glass and helmet), body-dressed (such as coat, underwear, and trousers), hand-worn (such as watch, bracelet, and gloves), and foot-worn (such as shoes and socks).

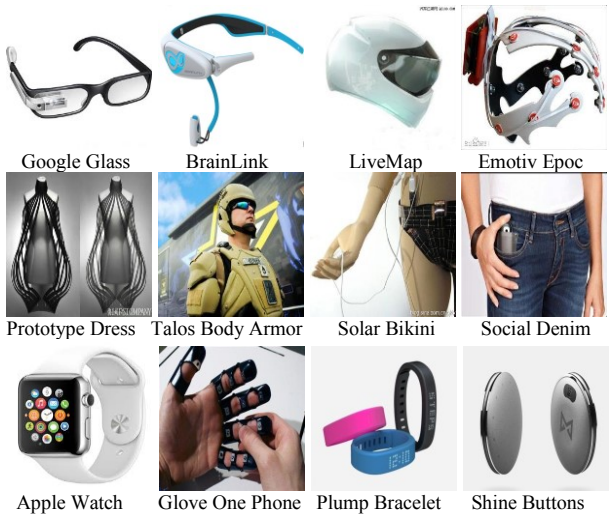

Figure 1. A variety of wearable devices

Another standard is based on product functions, including healthy living (such as sport wristband and smart bracelet), information consulting (such as smart glass and smart watch), and somatosensory control (such as somatosensory controller). Figure 1 lists a variety of wearable devices.

## II. WHAT MAKES WEARABLE DEVICES DIFFERENT?

A wearable device is more convenient for users to use and carry due to its miniaturization, lightweight and dressing. Their functions, forms and usages are different from tablet computers and mobile phones. In the wearable symposium held in 1997 [8], Bass L. summarized five characteristics that a wearable device should have [9]:

1) It may be used while the wearer is in motion;

2) It may be used while one or both hands are free or occupied with tasks;

3) It exists within the corporeal envelope of the user, i.e., it should be not merely attached to the body but becomes an integral part of the person's clothing;

4) It must allow the user to maintain control;

5) It must exhibit constancy, in the sense that it should be constantly available.

Steve Mann also provided the definition of wearable devices and described them from three operational modes and six attributes at the International Conference on Wearable Computing (ICWC) [1] held in 1998, The three operational modes include constancy, augmentation and mediation. And the Six attributes include unmonopolizing of the user's attention, unrestrictive to the user, observable by the user, controllable by the user, attentive to the environment and communicative to others.

However, wearable devices show more features as they evolve, such as diversity and concealment [10]. Wearable devices not only change the human-computer relationship and the way people use computers. Moreover, they produce a significant influence on people's life and work.

## III. SOFTWARE FOR WEARABLE DEVICES

The technology of wearable devices is not mature. The development of wearable devices is thus bound to encounter various problems such as functional singleness, incompatibility between operating system and software, convenience of human-computer interaction, data transmission, confidentiality of the information, energy consumption problems brought by continuous running. This chapter discusses some primary research issues in the development process of wearable devices.

### A. Operating System

Operating system is the interface of hardware and software. Its function is to manage hardware, software and data resources, to control program execution, to improve human-computer interaction, to enable users to have a good working environment, and to provide services for users and support for other applications.

The operating system on wearable devices has gone through years of development [11]. As early as in 2000, IBM collaborates with the famous Japanese watch manufacturer Citizen to launch a smart watch named WatchPad with Linux as its operating system. Fossil designed in 2003 a wrist device called Wrist PDA. It equipped with PalmOS [12] operating system and supported screen touch, which were very popular at that time. In addition, Microsoft designed in 2004 the SPOT system for smart watches. In 2013, Samsung released its first smart watch Galaxy Gear using Android as operating system. After that, Samsung launched the second generation of smart watch running Samsung's independently designed operating system Tizen. In March 2014, based on Android, Google launched a smart watch dedicated operating system called Android Wear [13], whose operation is implemented through Google Now's voice commands. Android Wear is expected to build a uniform and standard operating system platform, accelerating the development of wearable devices.

At present, there exist a variety of operating systems in wearable market. But they may not be convenient for users to use. Developers are difficult to choose which operating system for the device. And the application for one operating system is not suitable for another.

Since operating system is essential for wearable devices, we should design wearable operating systems by taking the features of wearable devices into account so that we can achieve the following objectives.

1) Convenience. The design of the operating system should be more convenient for users to use wearable devices.

2) Effectiveness. The operating system should be managed more effectively and take advantage of resources like hardware, software and data of wearable devices.

3) Scalability. The operating system should permit new system functions to be developed, tested and included.

4) Openness. The operating system should support integrated and collaborative network work of different manufacturers and devices so that it can achieve the portability and interoperability of applications.

5) Multitasking. The operating system should be able to run multiple applications concurrently.

In general, there are several options for developers to develop wearable operating system [14][15]:

1) Develop further on the basis of palm operating system used on mobile phones or other terminals. The advantage of this approach is to shorten the development cycle and to reuse existing applications. Its disadvantage is that there exist some problems on transplanting the existing applications.

2) Develop proprietary operating system based on Linux. This approach has strong pertinence. However, it is difficult for development and establishing the corresponding software of ecological environment.

3) Develop web-based operating system [14]. This approach would take full advantage of resources and support of the web server, adjust and restructure software system dynamically according to the actual requirements, minimize the resource requirements to wearable devices. The approach has great potential for development, but will take a long time and a large cost.

### B. Database Management System

Wearable devices run continuously, and are always ready to interact with users. They will collect user information as much as possible, and store them in database. The user can access the database to query necessary information via the database management system. For example, the Aid smart cane designed by Egle Ugintaite [7], utilizes built-in sensors to record user's pulse, blood pressure and temperature in real time, and displays health data through its LCD screen. The user can consult the database to obtain various data about his/her recent health status.

In addition to the basic information including gender, age, height and weight, wearable devices can also record user's health status, consumption habits, personal temper, and preference information to color and food, etc. These information may be useful for users achieving their goals. For example, when it is time for lunch, wearable devices are able to search nearby restaurants meeting the user's taste according to the database information, and provide hints to the user; when shopping, the devices can also recommend certain commodities that the user may need or may be interested in based on personal preference; when docking smart appliances, wearable devices can send personal information stored in the database to the terminals, and instruct smart appliance system to control air conditioning in real time according to the user's temperature, or automatically open the audio device and play music fitting the user's mood at that time.

At present, the major task of the database management system for wearable devices is only to deal with simple data, even some devices do not include the function of database management. The response speed and processing power may not necessarily meet the needs of devices.

Due to the different functionalities of wearable devices, the recorded data are changeful and may include the basic personal information, physical health data, and external environment changing data etc. To store these data, not only local databases, but also cloud databases are required and a reliable database management system becomes necessary. Therefore, developers should design specialized database management systems that are able to manage and operate various data, being lightweight and having fast response speed; or should transfer existing database management systems on mobile phones to wearable devices.

### C. Network Communication Protocol

Wearable devices may need to exchange data with other devices such as mobile phones, computers and other wearable devices. Therefore, network protocols for wearable devices become necessary. Network protocols define the network communication mode of wearable devices, and determine exchange data format and problems related to synchronization.

Bluetooth (IEEE 802.15.1), ZigBee (IEEE802.15.4), WiFi, NFC (IEEE 802.11) are four currently popular short-range wireless communication protocol standards [16]. In 2013, Broadcom launched Wireless Internet Connectivity for Embedded Devices (WICED) [17]. WICED simplifies the implementation of WiFi in a series of consumer electronics. It can achieve WiFi networking application or connect mobile phones, tablets computers and wearable devices for data sharing. Bluetooth Smart [18] is an improved version of Bluetooth 4.0 [19] with the power reduced by ten to twenty times. Furthermore, even if the radio is closed in most of the time, devices can still keep connected; when data are ready, devices can be awakened immediately within a very short time. In this way, battery life can be extended, which makes Bluetooth Smart more suitable for wearable devices.

At present, network communication protocols for wearable devices are relatively simple and focus primarily on wireless functions. However, wearable devices will eventually implement more and more functions like those implemented in mobile phones, such as WAP, GRS, GPRS, large file or data transmission, etc. Therefore, more reliable network communication protocol supports are demanded.

Developers should transfer network communication protocols running on tablet computers or mobile phones to wearable devices, or design special network communication protocols for wearable devices, which are more energy-efficient, safe, and with high throughput.

### D. Application Development Platform

The rapid development of mobile phones attributes the success to the large number of available applications. Similarly, the lack of applications is one of the obstacles for the rapid development of wearable devices. Software development requires development environments and a complete set of tools, including modeling tools, interface tools, project management, configuration management tools, and testing tools, etc. Software development kit is often used to support a particular software engineering method, reduce the burden of manual management, and make software more systematic like other software engineering methods.

At present, there are many hardware design platforms for wearable devices. In 2013, Broadcom launched a development platform called WICED [17]. Bluetooth, WiFi, NFC and positioning technology can be integrated into wearable devices. The wireless networking function with low power consumption, high performance and interoperability can also be embedded into devices. As the representative of the wearable system platform based on ARM architecture, Freescale launched the open-source, scalable Wearable Reference Platform (WaRP) [20] with over fifteen manufacturer collaborators including Kynetics, RevolutionRobotics, Circuitco. The platform supports open operating system such as Android and Linux and has features like scalability and flexibility.

Wearable market is currently in the stage of development and promotion of hardware, while application development is lagging behind. In order to accelerate the software application development of wearable market, various software reference platforms are springing up. In 2014, Google released the software development kit (SDK) including an emulator and other tools for Android Wear [21], which can allow developers to integrate Android Wear platform with their own devices and applications that can be available to download. Developers of the Android Wear ecological circle manufacturing can utilize all the tools they need to start making apps for the new devices. The development of SDK for Android Wear is expected to bring richer APP experience for future smart watches. Tizen is an open source operating system based on Linux for mobile phones and other smart devices. It has a complete development platform, including simulator, IDE etc. At present, Samsung's Galaxy Gear2 smart watch is running on this operating system. If the promotion is widely successful, we believe that Tizen will contribute to the development of wearable devices.

However, it is still lack of mature software development platform for wearable devices. An important issue for developers is to choose which platform to develop application software. In addition, developers cannot verify which applications have been developed and available. Therefore, developers have several options [22]:

1) Develop application software that supports particular operating systems for a single platform. This approach would simplify developers' work, but the resulting application may not be able to run on wearable devices with other different operating systems;

2) Develop native applications for each platform, but the technology and cost of application maintenance for each platform become the big challenges;

3) Develop mobile web applications, so as to reduce the native code for each platform. But the application may not be able to meet the market demand.

*E. Privacy and Security*

Wearable devices can collect real-time user information so as to provide effective personalized services for users. These data contain various user informations, for example, geographical location, living habits, body temperature, heart rate, account password and conversation, etc. When mishandled, it may bring great danger to the user's privacy and security, and harm the user's property or even personal safety. With the wide use of wearable devices and the rapid growth of applications, security and privacy issues should become more and more important. The communication of wearable devices is mainly based on wireless network. Thus, more private information may be easily attacked or stolen.

Wearable industry is not mature. There is still lack of a complete and effective program to protect privacy and security. While the industry may neglect security problems due to cost issues, researchers working in wireless sensors and mobile application have done a lot of fundamental work that will help to solve the privacy and security problems in wearable devices. For example, Liu et al. studied the security problems on mobile application [23]. Ameen et al. analyzed the problems of information security and privacy protection on wireless sensor network in the health care field [24].

We can consider from the following aspects to protect user's information and privacy:

1) Research on reliable network communication protocols for wearable devices so as to ensure the security of data during transmission;

2) The system should have permission setting. Wearable applications can only be allowed to obtain necessary data so as to reduce data exposure;

3) Reasonable software patterns should be proposed to solve these problems. For example, loosing the binding between private data and real names and mixing various may help to protect user's privacy and security.

*F. Energy Consumption*

Because wearable devices can only use battery, rather than stationary power, as their power supply, it is more tedious to charge wearable devices than mobile phones. Frequent charging or replacing battery can inevitably reduce the practicability of devices and the preference and satisfaction of users. In addition, the large amount of energy consumption for devices can produce great heat. If the cooling problem cannot be handled properly, it will damage user experience or even cause low temperature scald. Therefore, the energy consumption of wearable devices is an issue worthy of attention.

At present, manufacturers control energy consumption of wearable devices or mobile phones through the design of hardware or operating systems. Since the beginning of using wearable devices, their battery life is not very satisfactory. Taking smart watches as an example, the battery life of Moto 360 is up to 60 hours [25] and the one of LG G Watch is only 36 hours [26]. Energy consumption management is an essential issue for wearable devices.

In addition to hardware and operating systems, we can also consider improving energy consumption control from a high-level application layer.

Particularly, the control of energy consumption can be considered from the following aspects at the application layer.

1) Reduce hardware electricity consumption through reasonable invoke of system APIs. For example, Hao et al. and Li et al. proposed the code-level energy consumption analysis methods on mobile applications [27][28]. Such methods can be extended to wearable devices to achieve the purpose of reducing energy consumption through invoking less-energy-consumption API or arranging reasonable invoking sequence.

2) Create adaptive energy-sensitive applications to adjust automatically energy usage. When the energy is sufficient, high quality services will be provided; otherwise, unimportant applications will be turned off in order to increase the usage time. Mizouni R et al used in [29] such adaptive strategy to reduce the energy consumption of smart phones in mobile applications.

3) Adopt load-balancing method to transfer complex calculation to the mobile terminal via wireless communication network, thus reducing wearable device's own energy consumption. Kwon et al. presented a method to solve the problem [30]. A similar approach can also be introduced into wearable devices through replacing high calculation energy consumption with low communication energy consumption.

*G. Human-Computer Interaction*

A major feature of wearable devices is to collect user information, perceive user physical conditions and the change of external environment, complete various commands, and assist or remind the user automatically [14]. Wearable devices pursue people-oriented, namely requiring wearable devices being more suitable for the user. Wearable devices would be best to perceive, recognize and understand human emotion, and give intelligent, sensitive, friendly response. Human-machine interaction is a key technology of wearable systems, which should solve the interaction between users and wearable devices and improve the ability of environmental awareness. Therefore, advanced interaction means are a hotspot in the research of wearable devices.

There are many ways for users to interact with wearable devices [14]:

1) Contextual Awareness: Wearable devices continuously run and collect data, but in most cases, the user does not use them. Wearable devices should run independently, perceive the external environment, and transfer useful information to the user. Starner T et al. proposed in [31] visual environment perception method for wearable devices, and pointed out that a wearable device can observe its user to provide serendipitous information, manage interruptions and tasks, and predict future needs without being directly commanded by the user.

2) Augmented Reality: That is a technology that enhances users' awareness to real world through the information such as sound, video, graphics or GPS data generated by the computer [32]. The goal is to apply virtual information to the real world and to superimpose virtual object, scene or system message generated by computers to the real scene. Zhou et al. presented a design approach and a series of practical proposals of wearable user interfaces in real augmented environment [33].

3) Non-keyboard input: The most familiar way for people to input information into computers or mobile phones is through keyboard or mouse. However it is impossible to connect such input devices to wearable devices because of their miniaturization and lightweight. Users can interact with wearable devices through non-keyboard ways such as voice, handwriting, gestures, data glove etc. For the disabled, the interaction ways of traditional smart devices cannot bring normal experience. But they can wear a wearable device that receives messages from other sensors, which are transmitted to their sensory system after analysis. For example, for a person whose eardrum is broken, hearing devices are directly connected to his/her skull, which enables him/her to sense the voice that is not passing through ears.

Therefore, compared to computers and mobile phones, wearable devices can provide many different ways of human-computer interaction for users to strengthen their experience. But these ways cannot meet the demand of all users, developers should strengthen the study of human-computer interaction technologies:

1) Apply existing mature human-computer interaction ways to wearable devices, such as handwriting, voice and other non-keyboard input. These ways can be easily realized in wearable devices and better accepted by users.

2) Strengthen the research of currently-immature human-computer interaction ways, such as contextual awareness, augmented reality, mediated reality, etc. These ways can enhance user experience and make users have greater interests in wearable devices.

3) Propose new human-computer interaction ways. Some ways may be suitable for particular groups or particular environment. However, this approach will increase the burden to developers, and moreover, it may take time to study and practice these new ways.

*H. Software Engineering*

With the massive popularity of wearable devices, software engineering for wearable devices is becoming more and more important. Although the research of software engineering for wearable devices is still rare, we can predict some aspects/issues that may be of value and worth studying:

1) Demand analysis: how to perfect the requirement documents should be a problem worth studying. From users' comments on the application, the discussion in the BBS or the analysis of related online articles, users' expectation for the functionalities of the wearable devices can be acquired and thus be used to improve the requirement documents.

2) Code recommendation: many functions are likely to be reused in different applications. Through analyzing source code of existing applications, we can recommend function codes to developers to achieve fast development.

3) Application transplantation: it is not necessary to design different applications for each operating system. We can create API mapping among different platforms, and develop an ideal compiler that can complete the application development for multiple platforms.

Software engineering for wearable devices is an emerging field that provides many opportunities for researchers to investigate to.

*I. Big Data*

With the development of science and technology, we have entered the era of "big data" with "4V" [34], namely the volume, the velocity, the variety, and the veracity. Internet technology and mobile applications promote the development of the big data. Laurila et al. discussed the opportunities and challenges brought by big data in mobile applications [35].

Big data technology will play a great role in promoting the development of wearable industry. A large amount of data will be generated in using wearable devices, including basic personal information, health status, as well as the preference to food, cloth and color etc. We can gain a lot of useful information by using big data analysis technology, which will provide a great help for scientific research, social development and users' life. Big data technology can be used to analyze the huge amounts of data collected by wearable devices, and to detect user's body health factors. For example, the analysis for heart rate and blood pressure can understand user's body status and potential risks. It is an indisputable fact that big data technology can help wearable-device users enjoy more convenient life. Redmond et al. discussed the significance of big data technology for wearable devices in health care [36].

Conversely, wearable industry will also promote the development of big data. Compared to mobile phones, wearable devices will produce larger amount of data. So it needs big data technology to deal with these collected data even more. For example, the wearable market has emerged various smart bracelets with advanced functions, design and parameters; however, these wearable devices cannot use big data technology for processing these data to provide useful information to users.

Wearable devices and big data technology will complement each other and pursue a common development. In the big data environment, wearable market will have great development, create enormous wealth, and make people's life more convenient.

## IV. CONCLUSION

Wearable devices will become the mainstream of the development of mobile smart devices, and dramatically change modern way of life. Currently, the development of wearable devices is still in its immature stage, and the major functions focus on running calculation, navigation, remote picture and other related services. However, these services can also be achieved on smart phones. Meanwhile, research on hardware materials and battery life has not achieved a breakthrough; limited screen space makes the product design very difficult; application software development is still in the initial stage. Due to these problems, it will take a long time for wearable devices to become the mainstream of market.